\documentclass{ws-mplb}
\usepackage[super]{cite}
\usepackage{hyperref}
\usepackage{url}

\begin{document}

\markboth{Shuai Dong and Jun-Ming Liu}{Recent progress of multiferroic perovskite manganites}

%
\catchline{}{}{}{}{}
%

\title{RECENT PROGRESS OF MULTIFERROIC PEROVSKITE MANGANITES}

\author{SHUAI DONG}
\address{Department of Physics, Southeast University, Nanjing 211189, China\\
sdong@seu.edu.cn}

\author{JUN-MING LIU}
\address{Laboratory of Solid State Microstructure, Nanjing University, Nanjing 210093, China\\
Institute for Advanced Materials, South China Normal University, Guangzhou 510631, China\\
International Center for Materials Physics, Chinese Academy of Sciences, Shenyang 110016, China\\
liujm@nju.edu.cn}

\maketitle

\begin{history}
\received{(29 February 2012)}
\revised{(3 Jun 2012)}
\end{history}

\begin{abstract}
Many multiferroic materials, with various chemical compositions and crystal structures, have been discovered in the past years. Among these multiferroics, some perovskite manganites with ferroelectricity driven by magnetic orders are of particular interest. In these multiferroic perovskite manganites, not only their multiferroic properties are quite prominent, but also the involved physical mechanisms are very plenty and representative. In this Brief Review, we will introduce some recent theoretical and experimental progress on multiferroic manganites.
\end{abstract}

\keywords{Multiferroicity; manganites; spiral spin order; exchange striction; CaMn$_7$O$_{12}$.}


\section{Introduction}

Multiferroicity denotes the co-existence of more than one primary ferroic order parameter simultaneously in a single material.\cite{Schmid:Fe} These ferroic order parameters can be ferromagnetism, ferroelectricity, ferroelasticity, and  ferrotoroidicity.\cite{Aken:Nat} To date, the most studied multiferroics are those materials with ferromagnetism (or antiferromagnetism) and ferroelectricity.\cite{Wang:Ap}. Since ferromagnetism and ferroelectricity are very useful in spintronics, information storages, sensors, etc., multiferroics with spontaneous magnetization and polarization are promising candidates to design advanced devices with faster speeds, more functions, and energy saving.

However, in many materials, the magnetism and ferroelectricity are mutually exclusive,\cite{Hill:Jpcb} which once blocked the development of multiferroicity in the last century. Till 2003, two milestone works, the discovery of magnetic-field-controllable ferroelectric polarization in TbMnO$_3$ crystals and a giant ferroelectric polarization in BiFeO$_3$ films, revived the research interests on multiferroicity.\cite{Kimura:Nat,Wang:Sci} Since then, multiferroics have become a flourishing research area. More and more multiferroic materials have been discovered and the understanding of underlying physical mechanisms have also been pushed forward gradually.

Up to date, there are already many experimentally verified multiferroic materials, most of which are oxides although there are also some exceptions, e.g. some fluorides and organic tetrathiafulvalene-$p$-bromanil.\cite{Scott:Jpcm,Kagawa:Np} According to Khomskii's classification,\cite{Khomskii:Phy} some of them belong to the type-II multiferroics (or so-called magnetic multiferroics), in which the origins of ferroelectric polarizations are relevant to some particular magnetic profiles, while in the type-I multiferroics the origins of polarizations are almost independent of magnetism. Therefore, the type-II multiferroics have strong intrinsic magnetoelectric couplings, or more precisely, their ferroelectric polarizations can be significantly affected by tuning their magnetism, while in most type-I multiferroics, the magnetoelectric couplings are usually weak. Therefore, the type-II multiferroics are more interesting in physics, and will be also important in future applications.

\begin{figure}
\centering
\includegraphics[width=\textwidth]{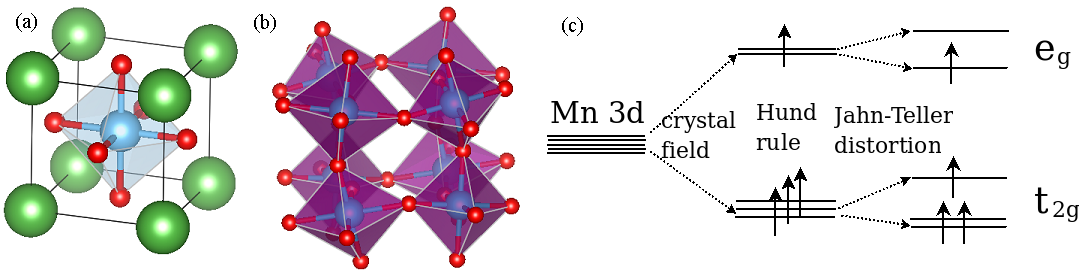}
\caption{(a) The perovskite structure. Center: B-site Mn cation; Corner: A-site rare earth/alkaline earth cations; The octahedral cage surrounding Mn is constructed by oxygen anions. (b) In many perovskite oxides, oxygen octahedra tilt and rotate collectively. These distortions reduce the nearest-neighbor hoppings (thus the bandwidth) of $e_{\rm g}$ electrons. (a) and (b) are drawn by VESTA.\cite{Momma:Jac} (c) Schematic drawing of split of Mn's $3d$ levels by the crystal field and Jahn-Teller distortions.}
\label{perovskite}
\end{figure}

In this Brief Review, multiferroic perovskite manganites will be introduced, which are typical type-II multiferroic materials. Among various type-II multiferroics, these perovskite manganites are the most important because not only their multiferroic properties are quite prominent, but also the involved physical mechanisms are very plenty. For example, the largest polarization of type-II multiferroics, always belongs to perovskite manganites so far, although this record has been updated year by year. In addition, almost all known physical mechanisms involved in the type-II multiferroicity, exist in perovskite manganites. In fact, most of these mechanisms were firstly revealed in perovskite manganites.

In the past years, there were already many excellent reviews on the multiferroicity.\cite{Wang:Ap,Fiebig:Jpd,Cheong:Nm,Ramesh:Nm,Tokura:Am,Picozzi:Jpcm,Kimura:Armr,Nan:Jap,Khomskii:Phy,Spaldin:Jpcm,Shuvaev:Jpcm} For this reason, the present Brief Review will select some recent theoretical and experimental progress on multiferroic perovskite manganites, especially those works after above reviews. Some key physical issues which have already been discussed in above reviews will also be updated according to the latest discoveries. Some unsolved debates will also be presented for further discussion.

\section{Multiferroic phases in $R$MnO$_3$}

\subsection{Origin of polarization in the spiral spin phase}

In the beginning, Kimura \textit{et al}. noticed that the ferroelectric polarization in TbMnO$_3$ relates to the incommensurate-commensurate (lock-in) transition at $28$ K, below which the magnetic modulation wave vector of Mn is locked at a constant value.\cite{Kimura:Nat} Then, by using the neutron diffraction technique, Kenzelmann \textit{et al}. discovered that the Mn's magnetic structure of TbMnO$_3$ below $28$ K is a transverse incommensurate spiral order, which breaks patial inversion symmetry and induces the magnetoelectricity.\cite{Kenzelmann:Prl}. Soon after, Arima \textit{et al}. confirmed the spiral order in Tb$_{1-x}$Dy$_x$MnO$_3$, which's spiral period shrinks with $x$.\cite{Arima:Prl}

Theoretically, in 2005, Katsura, Nagaosa, and Balatsky (KNB) first proposed a theory based on the spin supercurrent to explain the magnetoelectric effect in noncollinear magnets.\cite{Katsura:Prl} Accroding to the KNB theory, a noncollinear spin pair gives rise to a ferroelectric polarization in proportion to $\vec{e}_{ij}\times(\vec{S}_i\times\vec{S}_j)$, where $\vec{e}_{ij}$ is the unit vector pointing from site $i$ to site $j$, and $\vec{S}_i$ denotes the spin at site $i$. Later, Sergienko and Dagotto proposed that the (inverse) Dzyaloshinskii-Moriya interaction is responsible for the polarization in $R$MnO$_3$ ($R$ denotes a rare earth cation, here is Tb or Dy).\cite{Sergienko:Prb} The formula of Dzyaloshinskii-Moriya interaction is $\vec{D}_{ij}\cdot(\vec{S}_i\times\vec{S}_j)$ and $\vec{D}_{ij}$ is perpendicular to $\vec{e}_{ij}$.\cite{Dzyaloshinsky:Jpcs,Moriya:Pr} Besides, Mostovoy derived a similar expression for polarization induced by incommensurate spin density wave states (in fact the cycloid spiral spin states) from the phenomenological Ginzburg-Landau approach.\cite{Mostovoy:Prl06}. The KNB theory and Dzyaloshinskii-Moriya scenario seem to be equivalent formally, both of which depend on the spin-orbit coupling and are in proportion to the cross production of noncollinear spin pair ($\vec{S}_i\times\vec{S}_j$). These theories were further confirmed by neutron diffraction measurements on Gd$_{0.7}$Tb$_{0.3}$MnO$_3$ which established the correlation between the spin-helicity and electric polarization.\cite{Yamasaki:Prl}.

\begin{figure}
\centering
\includegraphics[width=\textwidth]{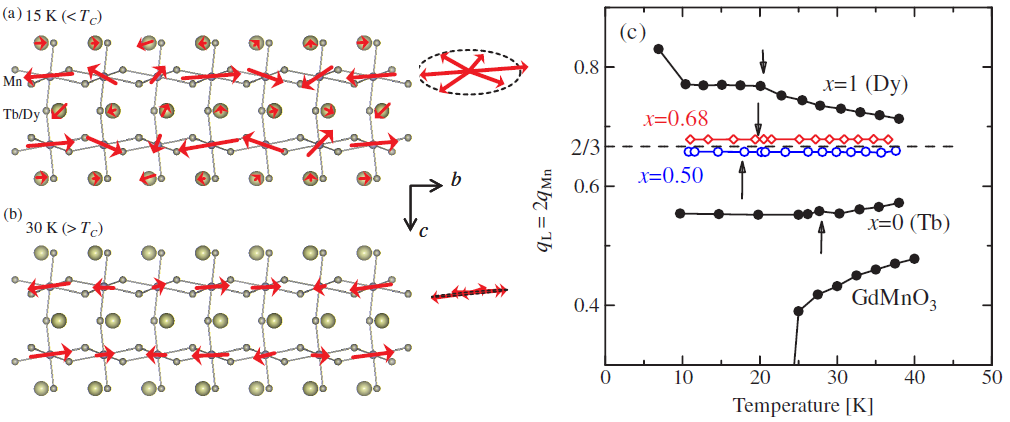}
\caption{Schematic drawing of the magnetic moments in Tb$_{0.41}$Dy$_{0.59}$MnO$_3$ projected onto the $bc$ plane at (a) $15$ and (b) $30$ K. The trajectories of the Mn spins are also shown. (c) Temperature dependence of lattice modulation wave number ($q_L$, which is twofold of the magnetic modulation wave number $q_{\rm Mn}$) in some Tb$_{1-x}$Dy$_x$MnO$_3$. The ferroelectric Curie temperatures are indicated by arrows. Reprinted figure with permission granted from \href{http://dx.doi.org/10.1103/PhysRevLett.96.097202}{T. Arima
\textit{et al}., \textit{Phys. Rev. Lett.} \textbf{96} (2006) 097202}.\cite{Arima:Prl}. Copyright \copyright (2006) by the American Physical Society.}
\label{spiral}
\end{figure}

However, in the KNB theory the polarization is purely electronic, while it is from ionic displacements in the scenario of Dzyaloshinskii-Moriya interaction. Since the (saturated) polarizations of $R$MnO$_3$ ($R$=Tb, Dy, Eu$_{1-x}$Y$_x$) are very weak (typically in the order of $0.1$ $\mu$C/cm$^2$, which is already quite prominent among magnetic multiferroics) comparing with traditional good ferroelectric materials, e.g. about $0.1\%$ of BiFeO$_3$ (in the order of $10^2$ $\mu$C/cm$^2$), it is not easy to direct measure the amplitude of ionic displacements in $R$MnO$_3$. Therefore, \textit{ab initio} calculation becomes an alternative approach to judge which one (electronic or ionic) is the dominant contribution in $R$MnO$_3$. The density functional theory calculation on TbMnO$_3$ by Malashevich and Vanderbilt found that the polarization from electronic contribution was very weak ($32$ $\mu$C/m$^2$, far below the real value $\sim600$ $\mu$C/m$^2$) and with an opposite sign against the KNB equation.\cite{Malashevich:Prl} Only when the ionic displacements were taken into account, a realistic polarization was reached in their calculation. Similar result was also obtained in Xiang \textit{et al.}'s density functional theory work.\cite{Xiang:Prl} In short, these theoretical calculations supported the ionic displacement scenario, at least for TbMnO$_3$. However, in a following work, Malashevich and Vanderbilt changed their tune a little because they found the proportions of electronic/ionic contribution seriously depended on the spiral plane ($ab$ or $bc$) as well as the octahedral rotations.\cite{Malashevich:Prb} In 2011, by using a diffraction technique which exploited the interference between the non-resonant magnetic scattering and the charge scattering arising from the ionic displacements, Walker \textit{et al}. measured the ionic displacements in TbMnO$_3$, which obtained very high precision values (up to several femto-meter) and decisively supported the microscopic models that attribute polarization to ionic displacements.\cite{Walker:Sci}

\subsection{Origin of the spiral spin order}

Above theoretical and experimental works have revealed how spiral magnetic orders generate ferroelectric polarizations. Then a following question is how to generate a spiral magnetic order? Or more specificly, why are there spiral spin orders in TbMnO$_3$ and DyMnO$_3$, instead of in LaMnO$_3$ or HoMnO$_3$. This question is quite nontrivial since the underlying physical mechanisms can guide us not only to tune the multiferroicity in manganites and related materials but also to search for new multiferroics with better performance.

\begin{figure}
\centering
\includegraphics[width=0.5\textwidth]{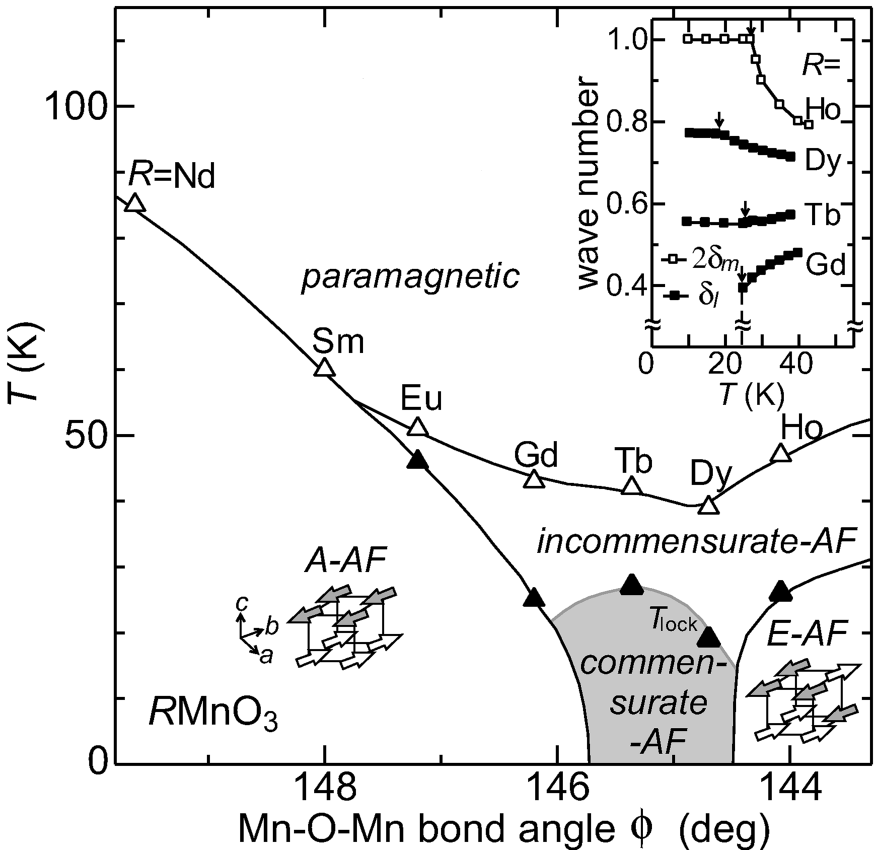}
\caption{Experimental magnetic phase diagram of $R$MnO$_3$. The horizontal axis stands for the decreasing Mn-O-Mn bond angle (increasing oxygen octahedra distortions) or namely the decreasing size of $R$. The N\'eel and lock-in transition temperatures are marked as open and closed triangles, respectively. Upper inset: Temperature dependencies of the wave numbers of lattice ($\delta_l$) or magnetic ($\delta_m$) modulation. The A-type and E-type antiferromagnetic structures are shown in their regions. The gray area is the spiral spin order state. Reprinted figure with permission granted from \href{http://dx.doi.org/10.1103/PhysRevLett.92.257201}{T. Goto
\textit{et al}., \textit{Phys. Rev. Lett.} \textbf{92} (2004) 257201}.\cite{Goto:Prl} Copyright \copyright (2004) by the American Physical Society.}
\label{RMnO}
\end{figure}

In doped manganites, there are many phases and the competition between these phases may give rise to dramatic responses to external stimulations, e.g. the colossal magnetoresistive effect and colossal electroresistive effect.\cite{Dagotto:Bok,Dagotto:Prp,Tokura:Rpp,Dong:Jpcm07,Dong:Prb07} In fact, even in undoped $R$MnO$_3$, there are also many competing phases. When $R$ is large (e.g. La and Pr) the magnetic ground state is A-type antiferromagnetic phase.\cite{Wollan:Pr} When $R$ is very small (e.g. Ho and Tm), it becomes E-type antiferromagnetic order.\cite{Munoz:Ic} In the middle region (e.g. Tb and Dy) between the A-type and E-type antiferromagnets, the ground state is the spiral spin order, as shown in the Fig.~\ref{RMnO}.\cite{Goto:Prl}

\begin{figure}
\centering
\includegraphics[width=0.5\textwidth]{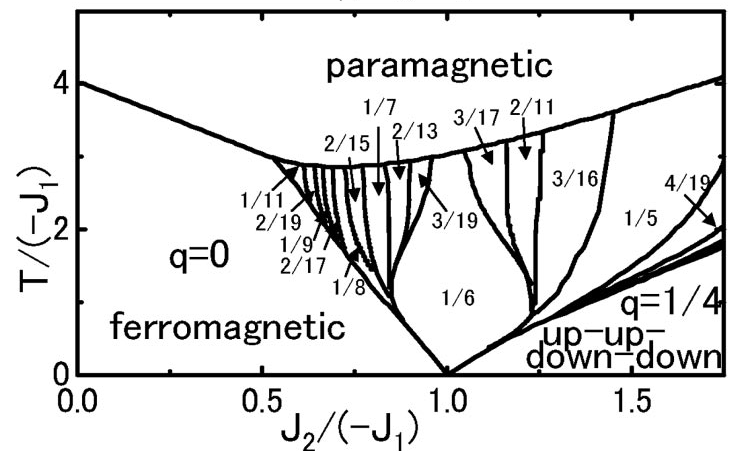}
\caption{Finite temperature (vertical axis) phase diagram of the two dimensional $J_1-J_2-J_3$ model. Here $J_3/J_1=0.01$. $q$ is the wave number of spin structure along the qseudocubic $x$ and $y$ axes. $q=0$ denotes the A-type antiferromagnetic order (ferromagnetic in the two dimensional lattice) and $q=1/4$ denotes the E-type states. The middle $q$'s stand for the incommensurate spin modulation. At that time, the spiral spin order had not been revealed. Reprinted figure with permission granted from \href{http://dx.doi.org/10.1103/PhysRevB.68.060403}{T. Kimura
\textit{et al}., \textit{Phys. Rev. B} \textbf{68} (2003) 060403(R)}.\cite{Kimura:Prb}. Copyright \copyright (2003) by the American Physical Society.}
\label{pd1}
\end{figure}

The first theoretical investigation was done as early as the discovery of multiferroicity in TbMnO$_3$. In 2003, Kimura \textit{et al}. studied a two dimensional frustrated classical spin Heisenberg model with ferromagnetic nearest-neighbor exchange ($J_1$), ferromagnetic/antiferromagnetic next-nearest-neighbor exchange along the $a$/$b$ axis ($J_3$/$J_2$) respectively.\cite{Kimura:Prb} The mean field phase diagram is shown in Fig.~\ref{pd1}. With increasing $|J_2|$, the magnetic transition occurs from the A-type antiferromagnet to the E-type antiferromagnet through the incommensurate structure.

However, this classical spin model needs a strong antiferromagnetic next-nearest-neighbor exchange $J_2$ to stabilize the spiral spin order and E-type antiferromagnetic order, like $J_2=-J_1$ to obtain the $q=1/6$ spiral. This abnormal large $J_2$ seems to be unreasonable in the orthorhombic lattice because the exchange path between next-nearest-neighbor sites is more complex and longer than the nearest-neighbor one. Furthermore, in this phase diagram, the incommensurate spin modulation fades away when temperature approaches zero, which disagrees with the experimental phase diagram.

\begin{figure}
\centering
\includegraphics[width=0.6\textwidth]{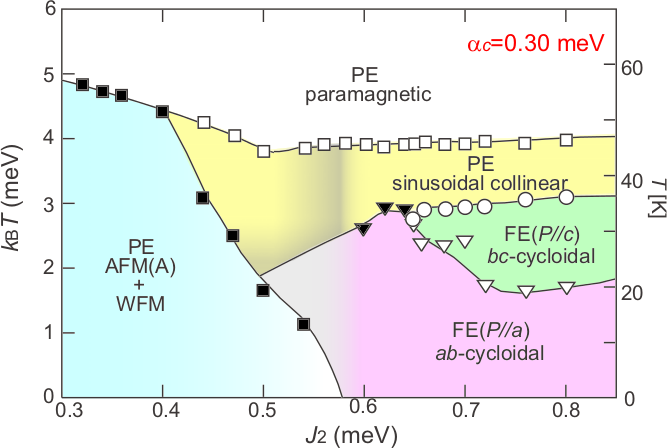}
\caption{A typical finite temperature magnetoelectric phase diagram of the Mochizuki-Furukawa model as a function of $J_2$. AFM͑ + WFM is the A-type antiferromagnetic phase with weak ferromagnetism due to the spin canting. PE/FE denotes paraelectric/ferroelectric phase, respectively. Reprinted
figure with permission granted from \href{http://dx.doi.org/10.1103/PhysRevB.80.134416}{M. Mochizuki \textit{et al.}, \textit{Phys. Rev. B} \textbf{80} (2009) 134416}.\cite{Mochizuki:Prb} Copyright \copyright (2009) by the American Physical Society.}
\label{mf}
\end{figure}

Mochizuki and Furukawa extended the frustrated classical spin Heisenberg model to the three dimensional lattice, including more interactions such as the single-ion anisotropy, Dzyaloshinskii-Moriya interaction, and cubic anisotropy.\cite{Mochizuki:Prb,Mochizuki:Jpsj} With proper parameters, the experimental phase diagram can be well reproduced, as shown in Fig.~\ref{mf}. The main conclusion is that the next-nearest-neighbor spin exchanges enhanced by the orthorhombic lattice distortion (the so-called GdFeO$_3$ type distortion) induces the spiral spin order, while the spiral plane ($ab$ \textit{vs}. $bc$) is exquisitely controlled by tuning the competition between the single-ion anisotropy and Dzyaloshinskii-Moriya interaction.

Using this model, Mochizuki \textit{et al}. studied many physical properties of multiferroic $R$MnO$_3$, such as the reorientation of polarization,\cite{Mochizuki:Prb,Mochizuki:Jpsj} electromagnons,\cite{Mochizuki:Prl} magnetostrictions,\cite{Mochizuki:Prl2} picosecond optical switching of spin chirality,\cite{Mochizuki:Prl3} magnetic switching of ferroelectricity.\cite{Mochizuki:Prl4} For more details of Mochizuki \textit{et al}.'s model, readers can refer to their recent long paper.\cite{Mochizuki:Prb11}.

Besides these models based on pure classical spins, there are also quantum models for $R$MnO$_3$, e.g. the double-exchange model.\cite{Hotta:Prl,Sergienko:Prb,Dong:Prb08.2,Dong:Epjb,Kumar:Prl10} In the double-exchange model, $e_{\rm g}$ electrons are treated as itinerant fermions, while $t_{\rm 2g}$ electrons are localized which's spin texture provides a magnetic background.\cite{Zener:Pr,Zener:Pr2,Goodenough:Pr} The double-exchange model has been extensively studied in the past decade to understand phase competitions and the colossal magnetoresistive effect in manganites bulks,\cite{Sen:Prl,Yu:Prb} as well as in thin films and heterostructures,\cite{Dong:Prb08,Dong:Prb08.3,Dong:Prb10,Dong:Prb11} which has been proved to be quite successful for manganites.\cite{Dagotto:Prp,Dagotto:Bok}

For undoped $R$MnO$_3$, both two $e_{\rm g}$ orbitals ($d_{x^2-y^2}$ and $d_{3z^2-r^2}$) must be considered in the double-exchange process. Beside the ferromagnetic double-exchange, there are antiferromagnetic superexchanges between neighbor $t_{\rm 2g}$ spins. In addition, the Jahn-Teller distortions also have to be taken into account.

\begin{figure}
\centering
\includegraphics[width=\textwidth]{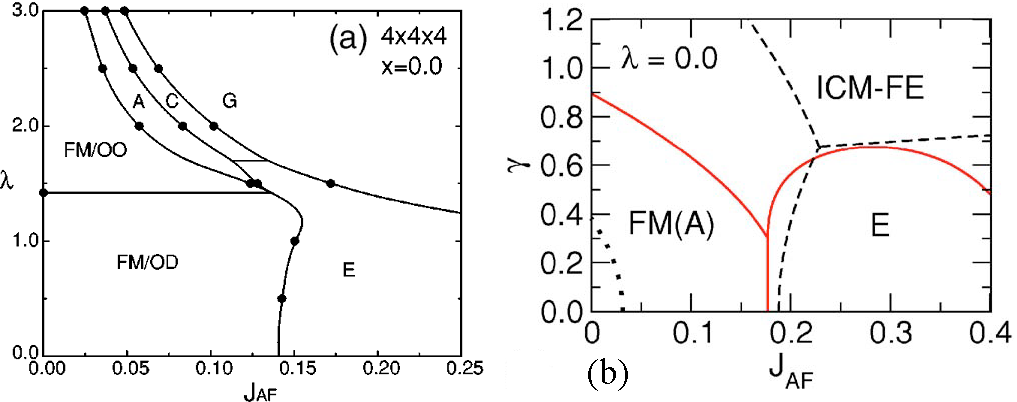}
\caption{Ground-state phase diagrams of the two-orbital double-exchange model for $R$MnO$_3$. The horizontal axis is the nearest-neighbor superexchange. (a) A mean-field result on a $4\times4\times4$ lattice. The vertical axis is the Jahn-Teller coefficient. FM denotes ferromagnetic order. OO (OD) denotes orbital order (disorder). A, C, E, G are various antiferromagnetic orders. Reprinted figure with permission granted from \href{http://dx.doi.org/10.1103/PhysRevLett.90.247203}{T. Hotta \textit{et al.}, \textit{Phys. Rev. Lett.} \textbf{90} (2003) 247203}.\cite{Hotta:Prl} Copyright \copyright (2003) by the American Physical Society. (b) With the Dzyaloshinskii-Moriya interaction (vertical axis). Dashed lines are the low temperature Monte Carlo results on a $8\times8$ lattice. Solid lines are calculated in the thermodynamic limit. Here the ICM-FE denotes the (incommensurate) spiral order phase with a ferroelectric polarization. Reprinted figure with permission
granted from \href{http://dx.doi.org/10.1103/PhysRevB.73.094434}{I. A. Sergienko \textit{et al.}, \textit{Phys. Rev. B} \textbf{73} (2006) 094434}.\cite{Sergienko:Prb} Copyright \copyright (2006) by the American Physical Society.}
\label{hotta}
\end{figure}

In an early study, Hotta \textit{et al}. investigated the phase diagram of $R$MnO$_3$ using this two-orbital double-exchange model.\cite{Hotta:Prl} However, only the A-type and E-type antiferromagnetic phases were reproduced, without the spiral one, as shown in Fig.~\ref{hotta}(a). Then, in Sergienko and Dagotto's work, the spiral spin order was obtained when the Dzyaloshinskii-Moriya interaction was included, as shown in Fig.~\ref{hotta}(b).\cite{Sergienko:Prb}. However, the required Dzyaloshinskii-Moriya interaction is too large, far beyond the real intensity.

\begin{figure}
\centering
\includegraphics[width=0.6\textwidth]{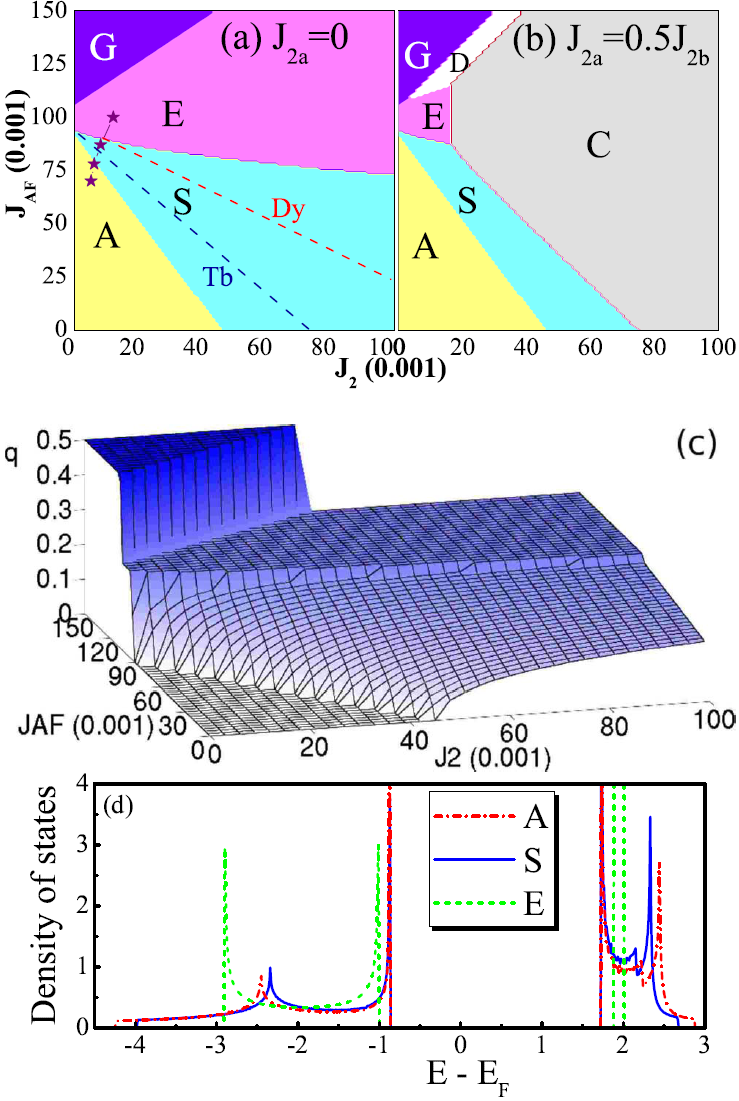}
\caption{Ground-state phase diagrams of two dimensional two-orbital double-exchange model for $R$MnO$_3$ with the next-nearest-neighbor superexchange $J_{2}$ and Jahn-Teller distortions. Vertical axis: the nearest-neighbor superexchange; Horizontal axis: the next-nearest-neighbor superexchange along the $b$-axis. A, S, E: A-type, spiral, E-type antiferromagnets, respectively. (a) $J_{2a}=0$. (b) $J_{2a}=0.5J_{2b}$. (c) The corresponding wave number (along the qseudocubic $x$ and $y$ axes) profile of magnetic phases in (b). (a-c) Reprinted figure with permission granted from \href{http://dx.doi.org/10.1103/PhysRevB.78.155121}{S. Dong \textit{et al.}, \textit{Phys. Rev. B} \textbf{78} (2008) 155121}.\cite{Dong:Prb08.2} Copyright \copyright (2008) by the American Physical Society. (d) The density of states for the A-type, spiral, E-type antiferromagnetic phases. Reprinted figure with permission granted from \href{http://dx.doi.org/10.1140/epjb/e2009-00225-1}{S. Dong \textit{et al.}, \textit{Eur. Phys. J. B} \textbf{71} (2009) 339}.\cite{Dong:Epjb} Copyright \copyright (2009) by EDP Sciences, Societa Italiana di Fisica, Springer-Verlag. Here the energy unit is the double-exchange hopping amplitude ($\sim0.2$ eV).}
\label{ase2}
\end{figure}

\begin{figure}
\centering
\includegraphics[width=0.6\textwidth]{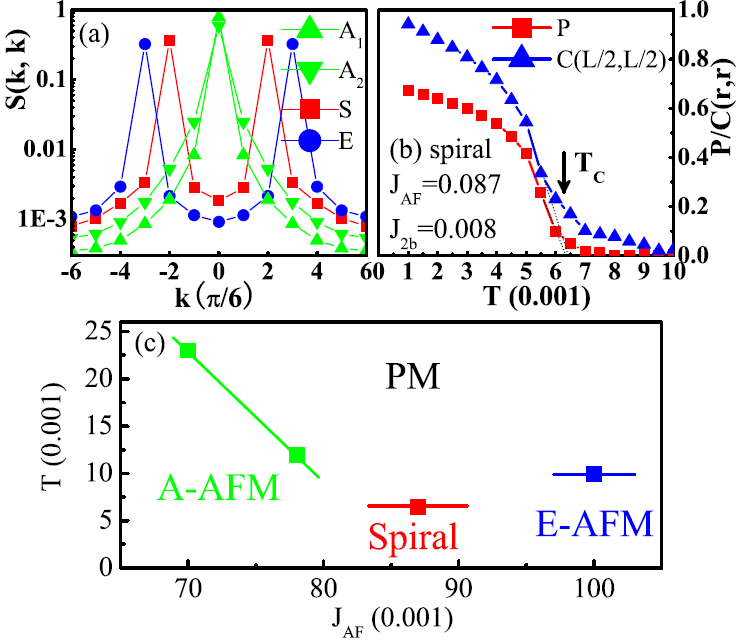}
\caption{(a) Typical spin structure factor of the four Monte Carlo simulations (corresponding to the four sets of parameters (asterisks) in Fig.~\ref{ase2}(a)) at a low temperature. (b) Temperature-dependent polarization ($P$) and long-range spin correlation ($C(L/2,L/2)$) corresponding to the spiral phase with a wave number $q=1/6$. All values are normalized to their saturation values. (c) Sketch of the finite-temperature phase diagram according to the transitions obtained in the Monte Carlo simulations. Reprinted figure with permission granted from \href{http://dx.doi.org/10.1103/PhysRevB.78.155121}{S. Dong \textit{et al.}, \textit{Phys. Rev. B} \textbf{78} (2008) 155121}.\cite{Dong:Prb08.2} Copyright \copyright (2008) by the American Physical Society.}
\label{ase4}
\end{figure}

To overcome these problems, Dong \textit{et al}. included the next-nearest-neighbor superexchange $J_2$ into the double-exchange model (without the Dzyaloshinskii-Moriya interaction since it is very weak).\cite{Dong:Prb08.2} Due to the GdFeO$_3$ type distortion of orthorhombic lattice, the next-nearest-neighbor exchange along the $b$-axis (in the $Pbnm$ notation) $J_{2b}$ is stronger than $J_{2a}$ along the $a$-axis. Dong \textit{et al}.'s calculation found that a weak $J_{2b}$ was enough to stabilize the spiral spin order. However, to obtain the realistic spiral periods, e.g. $\sim7.2$ cells for TbMnO$_3$ and $\sim5.2$ cells for DyMnO$_3$ (along the qseudocubic $x$ and $y$ axes), the contribution from Jahn-Teller distortions is essential. In these orthorhombic perovskites, the lattice constant along the $c$-axis shrinks, which corresponds to the Jahn-Teller $Q_3$ mode, and the in-plane ($ab$ plane) distortion (Jahn-Teller $Q_2$ mode) is also prominent. For undoped $R$MnO$_3$, these is an approximate ratio $|Q_2|\approx-\sqrt{3}Q_3$, which induces the special $d_{3x^2-r^2}$/$d_{3y^2-r^2}$ type of orbital order.\cite{Alonso:Ic,Zhou:Prb,Goodenough:Jmc} With proper Jahn-Teller distortions, a tiny $J_{2b}$, in the order of $10\%$ of the nearest-neighbor superexchange $J_{\rm AF}$, is enough to obtain the realistic spirals, as shown in Fig.~\ref{ase2}(a, b). Such a weak $J_{2b}$ and even weaker $J_{2a}$ seems much reasonable in the orthorhombic lattice.

With increasing $J_{\rm AF}$ and $J_{2b}$, the ground state changes from the A-type antiferromagnet to spiral phase, and finally to the E-type antiferromagnet, as found in real $R$MnO$_3$ with deceasing size of $R$. The following calculation on the three dimensional lattice also confirmed this A-S-E phase transition.\cite{Dong:Epjb}

The above zero-temperature phase diagrams should be further checked at finite temperatures. Monte Carlo simulations were employed to verify these spin orders and their phase transitions.\cite{Dong:Prb08.2} The results are shown in Fig.~\ref{ase4}. At low temperatures, all characteristic peaks of spin structure factor are prominent, suggesting robust spin orders. And the finite-temperature phase diagram qualitatively agrees with the experimental one.

Later, Kumar \textit{et al}. argued that even the weak $J_2$ was not necessary when using a finite Hund coupling (between the $t_{\rm 2g}$ spins and $e_{\rm g}$ spin).\cite{Kumar:Prl10} Their phase diagram was obtained by comparing zero-temperature energies of some candidate phases, which have not been further confirmed by unbiased Monte Carlo simulations. Other unexpected spin orders may emerge somewhere. In fact, the role of a weak $J_{2b}$ is not only to stabilize the non-collinear spin pairs, but also to make these cantings form uniform long-range chirality. Therefore, without this weak $J_{2b}$, a uniform spiral may be not stable against canting states with random local chirality.

\subsection{Multiferroicity of the E-type antiferromagnet}

In $R$MnO$_3$, when the size of rare earth cation $R$ is further reduced beyond Dy, the orthorhombic structure becomes unstable against the hexagonal structure. Even though, metastable orthorhombic structure remains available by using some special synthesis methods, e.g. under high-pressure conditions. In early days, experimental physicists were only interested in those spiral spin manganites like TbMnO$_3$ and DyMnO$_3$, while the E-type antiferromagnets (e.g. HoMnO$_3$) were rarely concerned.

\begin{figure}
\centering
\includegraphics[width=0.8\textwidth]{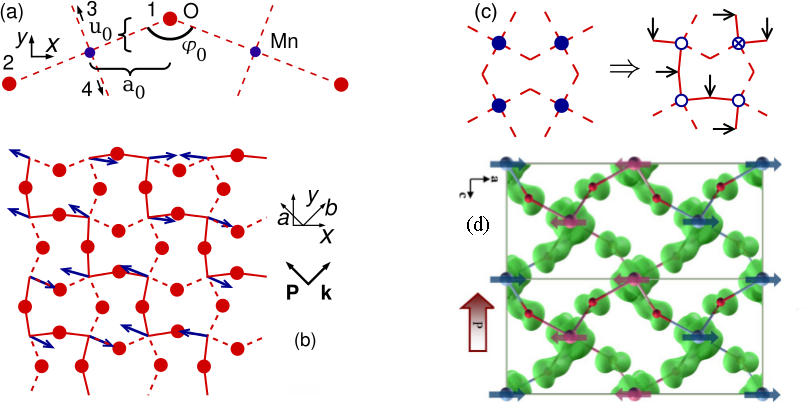}
\caption{Ferroelectric polarization in HoMnO$_3$. (a) The original (nonpolar) configuration of a Mn-O-Mn bond. $\varphi_0$ is the bond angle. (b) A Monte Carlo snapshot of the E-type antiferromagnetic phase. The ferromagnetic zigzag chain links are shown as solid lines. The displacements of the oxygen atoms are also exaggerated. (c) The local arrangement of the Mn-O bonds. Left: without any magnetic order; Right: with the E-type antiferromagnetic order. (a)-(c) Reprinted figure with permission granted from \href{http://dx.doi.org/10.1103/PhysRevLett.97.227204}{I. A. Sergienko \textit{et al.}, \textit{Phys. Rev. Lett.} \textbf{97} (2006) 227204}.\cite{Sergienko:Prl} Copyright \copyright (2006) by the American Physical Society. (d) A typical charge density isosurface plot of the MnO$_2$ plane. Arrows denote the direction of spins. Reprinted figure with permission granted from \href{http://dx.doi.org/10.1103/PhysRevLett.99.227201}{S. Picozzi \textit{et al.}, \textit{Phys. Rev. Lett.} \textbf{99} (2007) 227201}.\cite{Picozzi:Prl} Copyright \copyright (2007) by the American Physical Society.}
\label{E}
\end{figure}

In 2006, Sergienko, \c{S}en, and Dagotto predicted a large ferroelectric polarization driven by the spin zigzag chain in E-type antiferromagnets.\cite{Sergienko:Prl} In these systems, the mechanism for ferroelectricity is the exchange striction effect (relevant to $\vec{S}_i\cdot\vec{S}_j$), as shown in Fig.~\ref{E}, which is independent of the weak spin-orbit coupling. Therefore, the induced polarization in orthorhombic HoMnO$_3$ can be greatly enhanced comparing with those spiral magnetic multiferroics. The theoretical value of polarization was estimated to be $\sim2$ $\mu$C/cm$^2$ in orthorhombic HoMnO$_3$, $30$ times more than TbMnO$_3$.\cite{Sergienko:Prl}

The following density functional theory calculations further confirmed the giant ferroelectric polarization (up to $6$ $\mu$C/cm$^2$) in HoMnO$_3$.\cite{Picozzi:Prl,Yamauchi:Prb} The giant ferroelectric polarization in HoMnO$_3$ shows dual nature: a large portion arises due to quantum-mechanical effects of electron orbital polarization, in addition to the conventional polar atomic displacements.\cite{Picozzi:Prl}

However, the first measurement of polarizations of HoMnO$_3$ and its analogous YMnO$_3$ only found moderate values: $\sim90$ $\mu$C/m$^2$ for HoMnO$_3$ and $\sim 250$ $\mu$C/m$^2$ for YMnO$_3$, which were far below the theoretical predictions.\cite{Lorenz:Prb} Even considering the polycrystalline factor, the expected values of corresponding single crystalline samples were not very exciting.

The divergence between the theoretical predictions and experimental measurements has been gradually solved by improving samples' quality. Ishiwata \textit{et al}. synthesized a series of high quality polycrystalline samples of $R$MnO$_3$ ($R$=Eu$_{1-x}$Y$_x$, Y$_{1-y}$Lu$_y$, Dy, Ho, Er, Tm, and Yb), which showed a polarization up to $\sim800$ $\mu$C/m$^2$ for those E-type antiferromagnets.\cite{Ishiwata:Prb}  By multiplying a calibration factor ($=6$), the value for corresponding single crystals was estimated to be $\sim5000$ $\mu$C/m$^2$, as shown in Fig.~\ref{se}(a). which was much higher than the early results although still lower than the theoretical predictions. A possible reason is that the electric coercive forces of E-type antiferromagnetic multiferroics are very large, which require higher poling electric fields to obtain the saturated polarizations. By using a larger poling field ($37.5$ kV/cm$^{-1}$), Pomjakushin \textit{et al}. obtained a larger polarization up to $1500$ $\mu$C/m$^2$ in polycrystalline TmMnO$_3$, which seemed to be unsaturated yet.\cite{Pomjakushin:Njp} Recently, Ishiwata \textit{et al}. grew YMnO$_3$ single crystals under high pressure.\cite{Ishiwata:Jacs} The polarization with $10$ kV/cm$^{-1}$ poling field reached $2200$ $\mu$C/m$^2$ at $2$ K, which did not saturate either.\cite{Ishiwata:Jacs}

Comparing with bulks, thin films are more suitable to test higher poling fields. Nakamura \textit{et al}. fabricated a monodomain single-crystal film of orthorhombic YMnO$_3$ on the (0 1 0) YAlO$_3$ substrate.\cite{Nakamura:Apl} The measured in-plane polarization (with a poling field $10$ kV/cm$^{-1}$) reaches $8000$ $\mu$C/m$^2$ at $4$ K, which seems to be close to the saturation. The following resonant soft x-ray and hard x-ray diffraction revealed that the coexistence of E-type antiferromagnetic and cycloidal states below $35$ K.\cite{Wadati:Prl} The large polarization mainly comes from the E-type component in YMnO$_3$. Since this YMnO$_3$ thin film is not a pure E-type antiferromagnet, it is hopeful to expect a higher polarization in prototype E-type antiferromagnets at low temperatures.

Also recently, Lee \textit{et al}. synthesized rodlike large single crystals of orthorhombic HoMnO$_3$ utilizing the conventional Bi$_2$O$_3$ flux method.\cite{Lee:Prb} Unexpectedly, their HoMnO$_3$ showed an incommensurate magnetic modulation ($k\sim0.4$) at all temperatures,\cite{Lee:Prb} while an ideal E-type antiferromagnetic order should have a commensurate magnetic modulation ($k\sim0.5$) as found in previous polycrystalline samples.\cite{Munoz:Ic} As a result of this incommensurate magnetic modulation, the exchange striction within the Mn's zigzag chains does not contribute a net polarization any more. However, another exchange striction which is between the Mn's and Ho's spins, contributes a ferroelectric polarization along the $c$-axis, which reaches $1500$ $\mu$C/m$^2$ at $2$ K.\cite{Lee:Prb} Currently, it is unclear why their single crystals are so different from previous polycrystalline specimens. Lee \textit{et al}. proposed that the presence of defects or residual strains might play an important role. Further experiments on orthorhombic HoMnO$_3$ single crystals are needed to solve above puzzles and verify the real saturated polarization.\cite{Lee:Prb}

\subsection{Ferroelectricity driven by dual magnetic effects}

\begin{figure}
\centering
\includegraphics[width=0.7\textwidth]{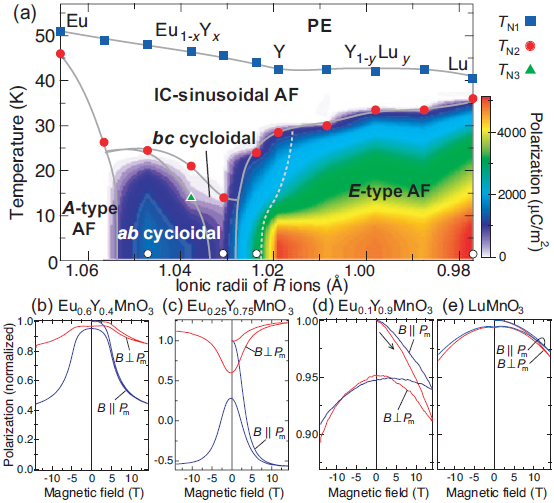}
\caption{(a) Contour plot of polarization in the phase diagram of orthorhombic $R$MnO$_3$ with nonmagnetic $R$ (Eu$_{1-x}$Y$_x$ and Y$_{1-y}$Lu$_y$). The polarizations in the E-type antiferromagnetic region are the polycrystalline values multiplying a calibration factor ($=6$). Magnetic-field dependence of normalized polarization at $2$ K for (b) Eu$_{0.6}$Y$_{0.4}$MnO$_3$, (c) Eu$_{0.25}$Y$_{0.75}$MnO$_3$, (d) Eu$_{0.1}$Y$_{0.9}$MnO$_3$, and (e) LuMnO$_3$. Reprinted figure with permission granted from \href{http://dx.doi.org/10.1103/PhysRevB.81.100411}{S. Ishiwata \textit{et al.}, \textit{Phys. Rev. B} \textbf{81} (2010) 100411(R)}.\cite{Ishiwata:Prb} Copyright \copyright (2010) by the American Physical Society.}
\label{se}
\end{figure}

In early studies, the Dzyaloshinskii-Moriya scenario (or the KNB theory) was applied to TbMnO$_3$ and DyMnO$_3$, while the exchange striction was applied to HoMnO$_3$ and YMnO$_3$. However, as shown in the above subsection, the latest experiments on YMnO$_3$ films and HoMnO$_3$ crystals showed some unusual results, which challenged the traditional viewpoint.\cite{Wadati:Prl,Lee:Prb}

According to Dong \textit{et al}.'s phase diagram, the phase transition between the spiral spin order and E-type antiferromagnetic order is first-order.\cite{Dong:Prb08.2} According to the experience in doped manganites with the colossal magnetoresistive effect, phase separations often emerge around first-order transition boundaries when some perturbations (e.g. quenching disorder) present.\cite{Dagotto:Prp,Dagotto:Bok,Tokura:Rpp}

Therefore, it is reasonable to expect the phase separation between the spiral spin order and E-type antiferromagnetic order in $R$MnO$_3$ by fine tuning some conditions, e.g. the average $R$ size. If so, the ferroelectricity can have dual origins: the Dzyaloshinskii-Moriya interaction effect as well as the exchange striction effect. The coupling between these two mechanisms may combine the giant ferroelectric polarization of exchange striction effect and the sensitive magnetoelectric response of Dzyaloshinskii-Moriya interaction effect in a single material.

In 2009, Lu \textit{et al}. synthesized Tb$_{1-x}$Ho$_{x}$MnO$_3$ polycrystalline samples, which's multiferroic transition temperatures showed a V-shape behavior with increasing $x$.\cite{Lu:Apa} This V-shape behavior is a common symptom for phase separation tendency in the colossal magnetoresistive manganites.\cite{Tokura:Rpp} Meanwhile, Ishiwata \textit{et al}. synthesized $R$MnO$_3$ ($R$=Eu$_{1-x}$Y$_x$ and Y$_{1-y}$Lu$_y$) polycrystalline samples, and also found the V-shape behavior between the spiral spin phase and E-type antiferromagnets, as shown in Fig.~\ref{se}(a). The magnetoelectric responses (sensitivity) are significantly enhanced in the phase boundary region: Eu$_{0.25}$Y$_{0.75}$MnO$_3$ (located at the bottom of V-shape valley in Fig.~\ref{se}(a)), as compared in Fig.~\ref{se}(b-e). This enhancement of magnetoelectric responses may arise from the phase separation between the spiral magnet and E-type antiferromagnet. As aforementioned, recent resonant soft x-ray and hard x-ray diffraction confirmed the coexistence of E-type antiferromagnetic and cycloidal states below $35$ K in YMnO$_3$ thin films.\cite{Wadati:Prl}

\begin{figure}
\centering
\includegraphics[width=0.6\textwidth]{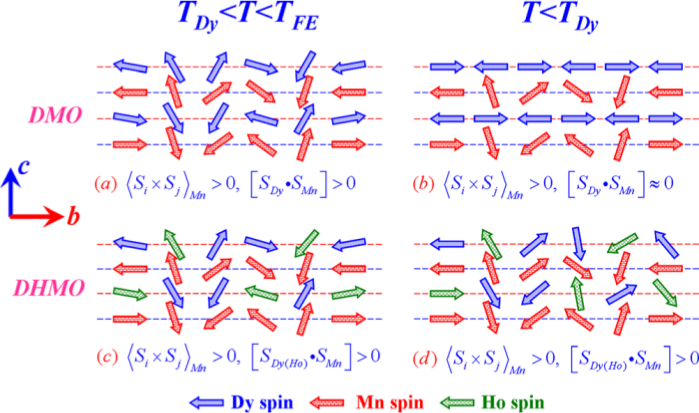}
\caption{Cartoon for the dual origins of ferroelectric polarization in Dy$_{1-y}$Ho$_y$MnO$_3$. The magnitude and orientation of these spin arrows are only for guide of eyes. Upper: pure DyMnO$_3$; Lower: Dy$_{1-y}$Ho$_y$MnO$_3$.  Left: within the temperature region [$T_{\rm Dy}$, $T_{\rm FE}$], where $T_{\rm FE}$ is the transition temperature of Mn's spiral spin order and $T_{\rm Dy}$ is the transition temperature of Dy's independent spin order. In this temperature region, Dy's magnetic moments form an incommensurate modulation following Mn's modulation period.\cite{Feyerherm:Prb,Feyerherm:Jpco} Right: at low temperatures ($<T_{\rm Dy}$). Both the Mn's spiral order and Mn-Dy(Ho)'s exchange striction contribute to the polarization in cases (a), (c), and (d). In case (b), only the Mn's spiral contributes the polarization since the independent Dy$^{3+}$ magnetic order cancels the exchange striction between Mn$^{3+}$ and Dy$^{3+}$. Reprinted figure with permission granted from \href{http://dx.doi.org/10.1063/1.3636399}{N. Zhang \textit{et al.}, \textit{Appl. Phys. Lett.} \textbf{99} (2011) 102509}.\cite{Zhang:Apl11.2} Copyright (2011) \copyright by the American Institute of Physics.}
\label{dhmo}
\end{figure}

Despite the coexistence of E-type antiferromagnetic and cycloidal states, the presence of A-site magnetism may also provide one more source of ferroelectric polarizations, e.g. in the forementioned HoMnO$_3$ single crystals.\cite{Lee:Prb} The exchange coupling between the A-site rare earth and B-site Mn can generate a polarization via the exchange striction effect. In fact, even in DyMnO$_3$, which was previously believed to be a prototype cycloidal spin multiferroic material, the exchange between Dy$^{3+}$ and Mn$^{3+}$ is responsible to the suppression of polarization at low temperatures. Furthermore, recent experiments on Dy$_{1-x}$Y$_{x}$MnO$_3$ and Dy$_{1-y}$Ho$_{y}$MnO$_3$ by Zhang \textit{et al}. revealed that the a considerable portion of polarization in DyMnO$_3$ is from the exchange striction effect between Dy$^{3+}$ and Mn$^{3+}$.\cite{Zhang:Apl11,Zhang:Apl11.2,Zhang:Fp} Then it becomes understandable why the improper polarization in DyMnO$_3$ ($\sim2500$ $\mu$C/m$^2$ under magnetic fields)\cite{Kimura:Prb05} is much larger than other spiral magnets (typically below $1000$ $\mu$C/m$^2$).\cite{Kimura:Armr} The exchange between $R$ and Mn also gives rise to prominent magnetoelectric responses in those E-type antiferromagnets with magnetic $R$, in contrast to the weak magnetoelectric responses in those E-type antiferromagnets with nonmagnetic $R$.\cite{Ishiwata:Prb} 

\begin{figure}
\centering
\includegraphics[width=0.6\textwidth]{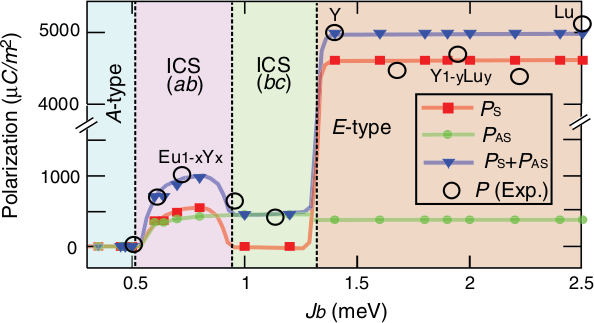}
\caption{(a) $J_b$ (the next-nearest-neighbor exchange) dependence of ground state polarizations. $P_S$ comes from the exchange striction effect while $P_{AS}$ comes from the Dzyaloshinskii-Moriya interaction. Both $P_S$ and $P_{AS}$ contribute to the polarization in the $ab$-spiral and E-type antiferromagnetic region. Reproduced from Ref.~Reprinted figure with permission granted from \href{http://dx.doi.org/10.1103/PhysRevB.84.144409}{M. Mochizuki
\textit{et al.}, \textit{Phys. Rev. B} \textbf{84} (2011) 144409}.\cite{Mochizuki:Prb11} Copyright \copyright (2011) by the American Physical Society.}
\label{dmes}
\end{figure}

In addition, there is one more case which can host the dual magnetic multiferroic effects. Mochizuki \textit{et al}. found the deformation of spin spiral can enable the exchange striction effect in those $ab$-plane spiral spin manganites, while in the $bc$-plane spiral spin manganites, this effect is canceled between different $ab$ layers.\cite{Mochizuki:Prl2,Mochizuki:Prb11}. Mochizuki \textit{et al}. also argued that there were noncollinear deformation of spins in the E-type antiferromagnets which was mostly believed to be collinear. If so, the Dzyaloshinskii-Moriya interaction would also contribute a small portion of polarization in the E-type antiferromagnets.\cite{Mochizuki:Prl2,Mochizuki:Prb11} However, this noncollinear deformation of spins needs further experimental verification.

\section{New multiferroic phases in doped manganites}

\subsection{Multiferroic SOS phase}

Despite their strong magnetoelectric coupling, the main weaknesses of $R$MnO$_3$ are the low multiferroic temperatures (most below $40$ K) and weak ferroelectric polarizations. Therefore, one of the most important issues for magnetic multiferroics is to improve the multiferroic performance, including both the working temperatures and polarizations.

In 2008, Kimura \textit{et al}. reported that CuO is a ``high''-temperature magnetic multiferroic material, which's multiferroic working temperature is $213-230$ K.\cite{Kimura:Nm} In 2010, Kitagawa \textit{et al}. reported that Z-type Sr$_3$Co$_2$Fe$_{24}$O$_{41}$ shows a spiral magnetic ordered state and a resultant magnetoelectric effect at room temperature.\cite{Kitagawa:Nm} In 2011, Lee \textit{et al}. reported that SmFeO$_3$'s magnetic multiferroic transition temperature reaches $\sim670$ K.\cite{Lee:Prl} However, in all these new multiferroics, the polarizations remain quite weak (typically in the order of $100$ $\mu$C/m$^2$ for single crystals).

Meanwhile, researchers are also searching in manganites for magnetic multiferroics with better performance. In 2009, by using density functional theory, Giovannetti \textit{et al}. predicted that half-doped manganites like La$_{0.5}$Ca$_{0.5}$MnO$_3$ could be multiferroic materials with very prominent polarizations up to several $\mu$C/cm$^2$.\cite{Giovannetti:Prl} This prediction based on an early idea of the site-centered \textit{vs}. bond-centered charge order in half-doped manganites.\cite{Efremov:Nm} However, till now, there is no direct experimental evidence to confirm the multiferroicity in half-doped manganites.

\begin{figure}
\centering
\includegraphics[width=0.5\textwidth]{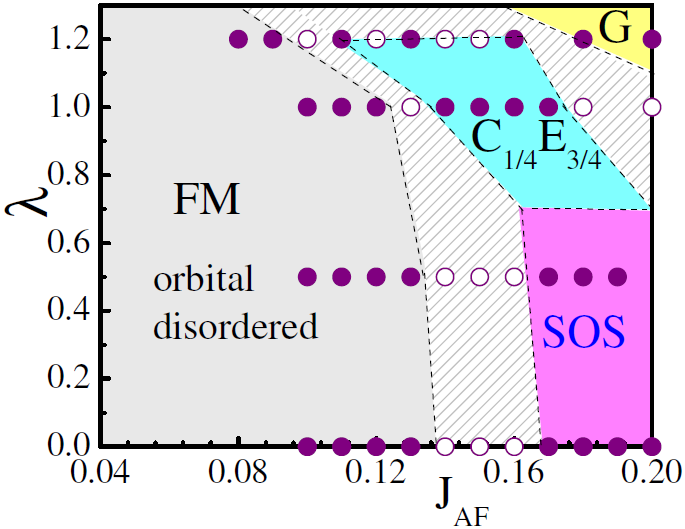}
\caption{Ground state phase diagram of quarter-doped two-orbital double-exchange model on the two-dimensional lattice. SOS is multiferroic. The horizontal axis $J_{\rm AF}$ is the nearest-neighbor superexchange and the vertical axis $\lambda$ is the Jahn-Teller coefficient. Full dots are those couplings where variational results have been confirmed by both the low-temperature Monte Carlo simulation and the zero-temperature relaxation on a $8\times8$ cluster. Open dots are cases where the Monte Carlo simulation does not provide a clear answer due to strong competition of metastable states. Reprinted figure with permission granted from \href{http://dx.doi.org/10.1103/PhysRevLett.103.107204}{S. Dong \textit{et al.}, \textit{Phys. Rev. Lett.} \textbf{103} (2009) 107204}.\cite{Dong:Prl} Copyright \copyright (2009) by the American Physical Society.}
\label{pdq}
\end{figure}

\begin{figure}
\centering
\includegraphics[width=0.6\textwidth]{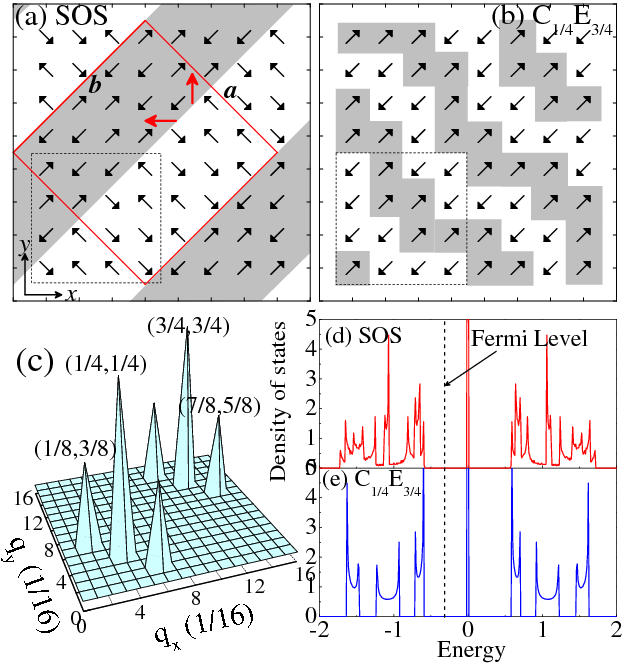}
\caption{Spin patterns of the (a) SOS and (b) C$_{1/4}$E$_{3/4}$ phases. (c) The (common) spin structure factor of these two phases. The characteristic peaks appear at ($1/4$, $1/4$), ($1/8$, $3/8$), and ($3/8$, $1/8$) and their equivalent positions in the $\vec{q}$ + ($1/2$, $1/2$) region. The ideal amplitudes of these three peaks are in the ratio of $2:1:1$. Density-of-states of the (d) SOS and (e) C$_{1/4}$E$_{3/4}$  phases (at Jahn-Teller coefficient $\lambda=0$, fixing the spin pattern). Both of these phases have an energy gap at the Fermi level, implying the insulating fact. Reprinted figure with permission granted from \href{http://dx.doi.org/10.1103/PhysRevLett.103.107204}{S. Dong \textit{et al.}, \textit{Phys. Rev. Lett.} \textbf{103} (2009) 107204}.\cite{Dong:Prl} Copyright \copyright (2009) by the American Physical Society.}
\label{sos}
\end{figure}

Also in 2009, Dong \textit{et al}. re-investigated the phase diagram of quarter-doped two-orbital double-exchange model for manganites.\cite{Dong:Prl} Although this phase diagram had once been studied by Hotta \textit{et al},\cite{Hotta:Prl01,Hotta:Prl} a new phase called SOS (for ``spin orthogonal stripe'' as explained below) was discovered by Dong \textit{et al}. in the narrow bandwidth but weak Jahn-Teller distortions region, as shown in Fig.~\ref{pdq}. In the narrow bandwidth and strong Jahn-Teller distortions region, it is the C$_{1/4}$E$_{3/4}$ phase predicted by Hotta \textit{et al}.\cite{Hotta:Prl01,Hotta:Prl} The spin patterns of SOS and C$_{1/4}$E$_{3/4}$ phases are shown in Fig.~\ref{sos}(a,b).

The C$_{1/4}$E$_{3/4}$ phase consists of zigzag chains of collinear spins, which can be viewed as a superstructure of $1/4$ C-type antiferromagnet and $3/4$ E-type antiferromagnet. It belongs to a general family as C$_{x}$E$_{1-x}$.\cite{Hotta:Prl,Hotta:Rpp} The most studied case of C$_{x}$E$_{1-x}$ is the CE phase (C$_{1/2}$E$_{1/2}$) in the half-doped manganites.\cite{Wollan:Pr,Dong:Prb}

Different from the C$_{1/4}$E$_{3/4}$ phase, the SOS order consists of stripe blocks. Within each stripe, spins are collinear, which form the E-type antiferromagnetic order. But between nearest-neighbor stripes, spins are orthogonal to each other. This noncollinear spin pattern is not caused by neither the next-nearest-neighbor exchange frustration nor the Dzyaloshinskii-Moriya interaction. Instead, it originates from the electronic self-construction at the particular doping concentration. Since the competing interactions involved in the SOS case are the double-exchange and superexchange, both of which are the nearest-neighbor interactions (thus both are strong), the phase transition temperature is expected to be higher than the spiral spin orders in undoped $R$MnO$_3$ which are stabilized by the weak next-nearest-neighbor interactions. According to the Monte Carlo simulation, the multiferroic $T_{\rm C}$ may reach $100$ K.\cite{Dong:Prl} The ferroelectric polarization of SOS phase has dual origins: the Dzyaloshinskii-Moriya interaction effect at the stripe boundaries and the exchange striction effect within every stripes. Thus, the expected polarization should be considerable.

Recently, Liang \textit{et al}. generalized the SOS phase to other doping concentrations $x=1/N$ ($N=2$, $3$, $4$, ... , $\infty$).\cite{Liang:Prb} A general SOS$_x$ family has been revealed, which corresponds the C$_{x}$E$_{1-x}$ family. The multiferroic phase predicted by Giovannetti \textit{et al}. can be included into this family as a special case SOS$_{1/2}$.\cite{Giovannetti:Prl} Some of these SOS$_x$ and C$_{x}$E$_{1-x}$ are multiferroic. More interestingly, there are various derivative superstructures of SOS$_x$ and C$_{x}$E$_{1-x}$, which are energy-degenerated with the corresponding primary ones. In other words, there is dimensional reduction in these complex spin structures.\cite{Liang:Prb}

\subsection{CaMn$_7$O$_{12}$}

Although there are many theoretical predictions for new multiferroics, few of them have been verified experimentally. Some predicted materials are very difficult to synthesize, or need very rigorous conditions. For example, to obtain the aforementioned SOS phase, one needs to get a quarter-doped perovskite manganites with very narrow $e_{\rm g}$ bandwidth but weak Jahn-Teller distortions.\cite{Dong:Prl} In normal perovskite manganites, the bandwidth decreases with decreasing size of A-site cations. However, the Jahn-Teller distortions will be enhanced simultaneously. Therefore, it seems almost impossible to simultaneously fulfill both conditions in normal perovskite manganites. Furthermore, the quenching disorder of A-site cations will seriously suppress the fragile long-range SOS order. This disorder problem does not exist in undoped $R$MnO$_3$. However, in doped manganites, the size and valence differences between the rare-earth cations and alkaline-earth cations are crucial to determine physical properties.\cite{Tokura:Rpp} Sr$^{2+}$ is certainly too large to reduce the bandwidth, while there is no size-matched rare-earth cation for Ca$^{2+}$.

\begin{figure}
\centering
\includegraphics[width=\textwidth]{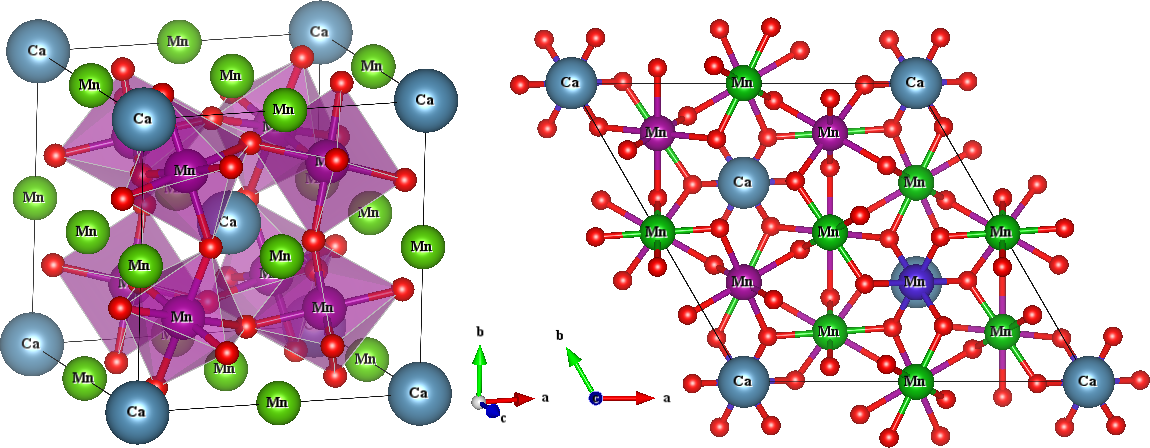}
\caption{Crystal structures of quadruple perovskite CaMn$_7$O$_{12}$. Left: cubic $Im\bar{3}$ at high temperatures ($>440$ K). Right: rhombohedral $R\bar{3}$ at low temperatures ($<440$ K). Green Mn's are at the A-site. Drawn by VESTA.\cite{Momma:Jac}}
\label{cmo}
\end{figure}

CaMn$_7$O$_{12}$, a quadruple perovskite, provides a unique structure to realize complex spin structures, e.g. the SOS phase. As shown in Fig.~\ref{cmo}, its lattice is cubic at high temperatures, with four perovskite units in each cell. This cubic cell distorts into a rhombohedral lattice at $\sim440$ K, accompanying a charge-order transition.\cite{Przenioslo:Jpcm} In CaMn$_7$O$_{12}$, Ca$^{2+}$ and three Mn$^{3+}$ occupy the A-site, while the rest four Mn's at the B-site have an average valence $+3.25$, which just corresponds to the quarter-doping. Comparing with normal perovskite manganites, CaMn$_7$O$_{12}$'s lattice size is greatly shrunk and the B-site Mn-O-Mn bonds are seriously distorted due to the small A-site Ca$^{2+}$ and Mn$^{3+}$, as required to reduce the $e_{\rm g}$ bandwidth. Furthermore, the Jahn-Teller distortions are very weak in CaMn$_7$O$_{12}$,\cite{Przenioslo:Pb} as also required to stabilize the SOS phase. The last but not the least, the A-site cations are fully ordered in quadruple perovskites, which eliminates the quenching disorder in normal doped perovskites. However, in the past years, CaMn$_7$O$_{12}$ have not been much studied and no direct evidence of multiferroicity has been reported.

\begin{figure}
\centering
\includegraphics[width=0.6\textwidth]{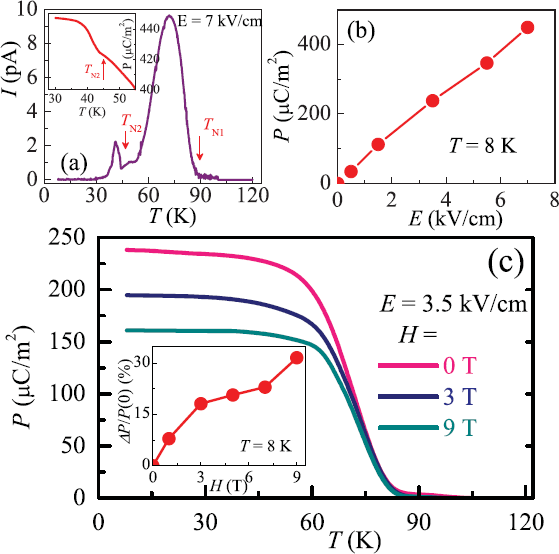}
\caption{(a) The pyroelectric currents of CaMn$_7$O$_{12}$ with a poling field $E=7$ kV/cm. An anomalous contribution appears around $T_{\rm N2}=48$ K (another magnetic transition temperature which's detail is unclear). Inset: The corresponding ferroelectric polarization around $T_{\rm N2}$. (b) The ferroelectric polarization at $8$ K as a function of the poling field $E$, which does not saturate up to $7$ kV/cm. (c) The suppression of polarization $P(H)$ by a magnetic field $H$. Inset: the magnetoelectric response ratio, defined as $[P(0)-P(H)]/P(0)\times100\%$. Reprinted figure with permission granted from \href{http://dx.doi.org/10.1103/PhysRevB.84.174413}{G. Zhang \textit{et al.}, \textit{Phys. Rev. B} \textbf{84} (2011) 174413}.\cite{Zhang:Prb11} Copyright \copyright (2011) by the American Physical Society.}
\label{zhang}
\end{figure}

In 2011, Zhang \textit{et al}. synthesized polycrystalline samples of CaMn$_7$O$_{12}$, and measured the ferroelectric polarization and magnetoelectric responses.\cite{Zhang:Prb11} As shown in Fig.~\ref{zhang}, the ferroelectricity is quite remarkable. First, the ferroelectric transition temperature is much higher comparing with other magnetic multiferroic manganites: $T_{\rm C}=90$ K, which coincides with the N\'eel transition $T_{\rm N1}$. Second, the polarization reaches $\sim450$ $\mu$C/m$^2$ at $8$ K when using a poling field $7$ kV/cm, which seems to be unsaturated yet. Third, this polarization can be significantly suppressed by $\sim30\%$ under a $9$ T magnetic field, which also suggests the magnetic origin of polarization.

\begin{figure}
\centering
\includegraphics[width=0.9\textwidth]{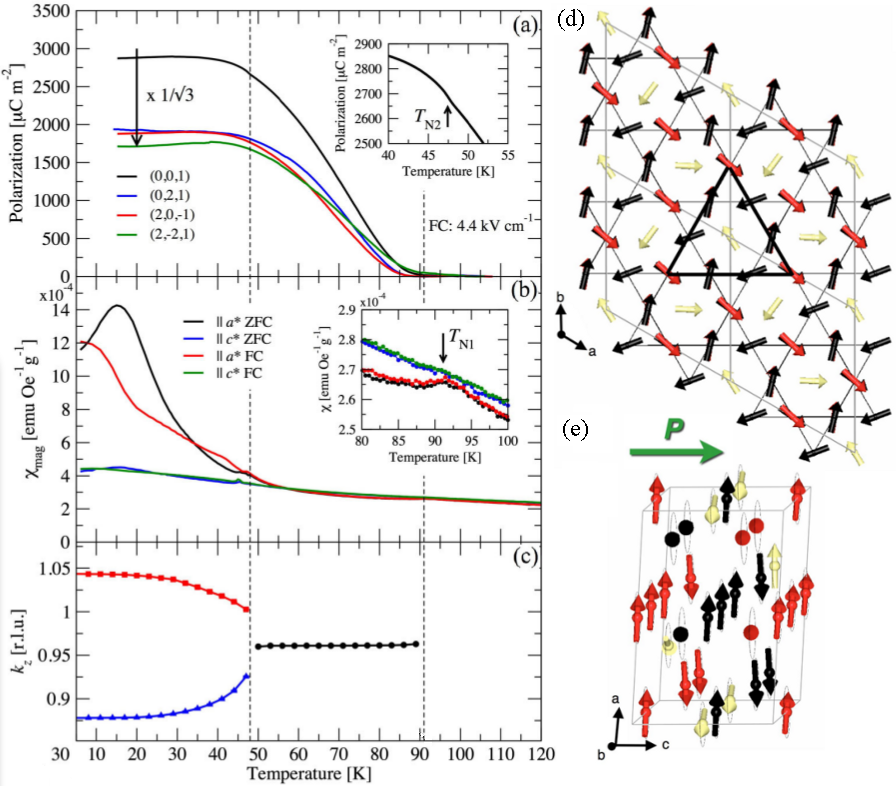}
\caption{(a) Polarization of CaMn$_7$O$_{12}$ single crystal along the hexagonal $c$ axis (black) and the three qseudocubic $<100>$ axes. Inset: A small anomalous at $T_{\rm N2}$. (b) The magnetic susceptibility parallel and perpendicular to the hexagonal $c$ axis, measured under a $500$ Oe magnetic field, in both zero field cooled and field cooled ($500$ Oe) conditions. Inset: the in-plane magnetic susceptibility at $T_{\rm N1}$. (c) Temperature dependence of incommensurate magnetic propagation along the $c$-axis. (d) and (e) are the magnetic structure of CaMn$_7$O$_{12}$ in the $ab$ and $ac$ planes, respectively. Mn1, Mn2, and Mn3 are shown in red, black, and yellow, respectively. The moments rotate in the $ab$ plane with a circular envelope as depicted in (e). Reprinted figure with permission granted from \href{http://dx.doi.org/10.1103/PhysRevLett.108.067201}{R. D. Johnson \textit{et al.}, \textit{Phys. Rev. Lett.} \textbf{108} (2012) 067201}.\cite{Johnson:Prl} Copyright \copyright (2012) by the American Physical Society.}
\label{johnson}
\end{figure}

Later, Jahnson \textit{et al}. synthesized single crystal samples of CaMn$_7$O$_{12}$ and determine the magnetic structures using the neutron powder diffraction.\cite{Johnson:Prl} The polarization, along the $c$-axis of the rhombohedral cell (the $[111]$ qseudocubic axis), reaches $2870$ $\mu$C/m$^2$ at low temperatures, which is one of the largest measured polarizations in magnetic multiferroics, as shown in Fig.~\ref{johnson}(a). Most importantly, the magnetic structure has been resolved. As shown in Fig.~\ref{johnson}(c), the magnetic wave vector is ($0$, $1$, $k$), where $k$ is $\sim0.963$ between $90$ K and $48$ K but splits into two branches below $48$ K ($\sim0.88$ and $\sim1.042$ at $5$ K). The in-plane and out-of plane spin patterns are shown in Fig.~\ref{johnson}(d,e). The incommensurate wave number $k$ suggests a screw type spiral propagating along the $c$-axis, which is different from the cycloidal spiral in undoped $R$MnO$_3$. Within the $ab$ plane, spins are also noncollinear.

However, the complex spin order in CaMn$_7$O$_{12}$ is not the expected SOS phase, which may be due to the magnetic interaction between the A-site Mn and B-site Mn. In the aforementioned model calculation, the A-site magnetism have not been taken into consideration, which needs further theoretical investigations.

Then, how to understand the giant improper ferroelectricity in CaMn$_7$O$_{12}$? According to the formula of Dzyaloshinskii-Moriya interaction (or the KNB theory), the screw type spiral along the $c$-axis can not give rise to a polarization, as in the Y-type hexaferrite Ba$_{0.5}$Sr$_{1.5}$Zn$_2$Fe$_{12}$O$_{22}$.\cite{Ishiwata:Sci} Johnson \textit{et al}. proposed that the ferroaxial coupling mechanism is responsible to the polarization. The symmetry group, $R\bar{3}$, belongs to the $\bar{3}$ ferroaxial class, for which a homogeneous structural rotation exists with respect to the high-temperature cubic group, represented by an axial vector $\vec{A}$. Symmetry considerations allow a possible coupling between the axial vector $\vec{A}$, the spin chirality $\sigma$ ($=\vec{e}_{ij}\cdot(\vec{S}_i\times\vec{S}_j$), which is a time-even, parity-odd pseudoscalar) along the $c$-axis, and the polarization $\vec{P}_c$ along the $c$-axis. This $\vec{P}_c\sigma\vec{A}$ term in free energy will give rise to a finite $\vec{P}_z$ when $\sigma$ is finite.\cite{Johnson:Prl,Mostovoy:Phy}

However, a recent density functional theory calculation showed that the screw spiral along $c$-axis is not crucial to obtain the giant ferroelectric polarization in CaMn$_7$O$_{12}$.\cite{Lu:ArX} Even with a zero $\sigma$, the in-plane noncollinear spin pattern can induce a polarization up to $\sim4500$ $\mu$C/m$^2$ along the $c$ axis. The underlying mechanism are the combination of Dzyaloshinskii-Moriya interaction and exchange striction.\cite{Lu:Prl} Therefore, further theoretical and experimental studies are needed to clarify the multiferroic mechanisms involved in CaMn$_7$O$_{12}$.

\section{Summary and perspective}

In this Brief Review, we have introduced some recent theoretical and experimental progress on multiferroic perovskite manganites, in which the ferroelectric polarizations are induced by particular magnetic orders. This review have (partially) emphasized three fundamental questions: 1) Why and how can the spiral spin order and E-type antiferromagnetic order generate improper ferroelectric polarizations? 2) Why are there the spiral spin order and E-type antiferromagnetic order in perovskite manganites and how to tune the competition between them? 3) Are there new magnetic multiferroics in doped manganites with better performance and how to understand their exotic properties?

To answer these questions, we have taken the undoped $R$MnO$_3$ as the example. The Dzyaloshinskii-Moriya scenario (or the Katsura-Nagaosa-Balatsky theory) and exchange striction effect have been applied to explain the origin of ferroelectricity. In some cases, these two mechanisms coexist and are mutually coupled. Both the classical spin model and quantum double-exchange model have been introduced to understand the phase diagram of $R$MnO$_3$. The theoretical SOS phase and experimental CaMn$_7$O$_{12}$ have also been introduced as new magnetic multiferroics in doped manganites.

The research field of multiferroicity remains very active and are developing very fast. The predictions/discoveries of new multiferroics have never been stopped since 2003. Accompanying with the progress on new multiferroic materials and improved samples' quality, the multiferroic performance has been improved gradually, including both the working temperatures and ferroelectric polarizations. In addition, the theoretical understanding of multiferroicity has also been pushed forward in the past years. Even though, there remains many puzzles and debates, some of which have also been presented in this review. Hope future experimental and theoretical works will solve them.

\section*{Acknowledgments}
This work was supported by the National Natural Science Foundation of China (Grant Nos. 50832002, 11004027, and 11074113), the National Key Projects for Basic Research of China (Grant Nos. 2011CB922101 and 2009CB929501) and NCET (Grant No. 10-0325).

\bibliographystyle{ws-mplb}
\bibliography{../ref}{}

\end{document}